\documentclass[a4paper,11pt]{article}
\usepackage{pos}
\usepackage{subcaption}
\usepackage{hyperref}
\usepackage{placeins}      %
\usepackage[sort&compress]{natbib}
\title{SS433  PeV neutron jet feeding the far TeV gamma beam}
\ShortTitle{SS433 PeV jet into TeV far beam}

\author*[\orcid{0000-0003-3146-3932}, a,b]{D. Fargion}
\author[\orcid{0000-0001-7503-2064}, c]{P.G. {De Sanctis Lucentini}}
\author[\orcid{0000-0002-4603-8405}, d]{S. Turriziani}
\author[\orcid{0000-0002-1653-6964}, e]{M.Y. Khlopov}
\author[e]{D. Sopin}
\affiliation[a]{Rome University “La Sapienza” and MIFP, Rome, Italy.}
\affiliation[b]{Osservatorio Astronomico di Capodimonte, INAF, Naples, Italy.}
\affiliation[c]{Physics Department, National University of Oil and Gas «Gubkin University», Moscow, Russia.}
\affiliation[d]{Centro de Astronomía (CITEVA), Universidad de Antofagasta, Av. Angamos 601, Antofagasta, Chile.}
\affiliation[e]{Virtual Institute of Astroparticle physics, 75018, Paris, France.}

\emailAdd{daniele.fargion@fondazione.uniroma1.it}

\abstract{
The SS433 is a well-known binary system with an internal black hole, which is stripping mass from an orbiting companion of ten solar masses, at a hundred of light-seconds away.
The black hole and its accretion disk fuel a thin precessing jet, whose spirals are well-observed.
Surprisingly, disconnected gamma-ray tails have recently been discovered by H.E.S.S., HAWC and LHAASO, hundreds of light-years away and with energies of tens of TeV.
We suggest that tens PeV neutron burst jets were ejected from the SS433 system over the past century.
These beams of ultra high-energy PeVatron neutrons, by their in-flight beta decay and Inverse Compton scattering,  could be the source of the enigmatic, distant and disconnected tens of TeV gamma-ray beams.
These ultra-relativistic PeV neutron jets could have been formed during one of the system's rare and intense tidal eruptions, when tens of PeV protons collide CV October 2025 with thermal ultraviolet photons, creating delta resonances. Their decay into secondary neutron beams of tens of PeV is well consistent with observations. Alternative models appear uncompetitive.
}

\ConferenceLogo{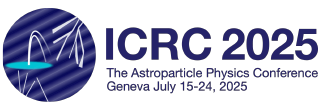}
\FullConference{39th International Cosmic Ray Conference (ICRC2025)\\
 15–24 July 2025\\
Geneva, Switzerland\\}
\graphicspath{{Figs/}{./}}
\def\orcid#1{\kern .08em\href{https://orcid.org/#1}{\includegraphics[keepaspectratio,width=0.6em]{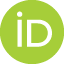}}}
\begin{document}
\maketitle

\section{Introduction}
Understanding microquasars is one of the frontiers of high-energy astrophysics. Their leading models are based on the capture and dismemberment of a star by a compact object, a neutron star (NS) or more generally a black hole (BH), while in a close orbit.
The mass of the companion star typically feeds an accretion disk around the NS or BH. The infalling mass also
feeds a precessing X-ray and gamma-ray jets, orthogonal to the same disk.
The spiral tail of the SS433 jets is the consequence of the precession of an ultra-relativistic outflow, spraying nucleons and electrons at relativistic speeds. The twin up-down jet has been observed in radio, X-ray, and gamma-ray spectra for nearly half a century. Its long spiral tails are diffuse and diluted over a distance of a light-year (ly). %
The source SS433 lies well within the supernova remnant nebula W50, whose small asymmetry reflects the jet's past and present role. Very recently, H.E.S.S.~\cite{hess2024acceleration} and HAWC~\cite{alfaro2024spectral} surprisingly discovered the resurgence of a twin hard gamma-ray signature, very distant and disconnected from SS433's inner jet. At a distance of about 75 or even 150 light-years from the same inner SS433 source.
A recent model is trying to explain the observations, based on an accelerating shock wave that will re-accelerate up to PeV energies the beam of nuclei and their TeV secondary tracks.
A re-collimation of the PeV-TeV beam jet is difficult to accept in the frame of this model.
We have considered and suggested an alternative model based on known high-energy nuclear physics.
Namely on a proton jet of tens of PeV converted by photon-pion scattering, within a past hot nova-like flare, into a neutron beam of tens of PeV \cite{bernabei2025proceedings,fargion2024ss433microquasarjettev}.  
This neutron jet, through its in-flight decay, is capable of simultaneously re-illuminating secondary electrons of tens of TeV and their inverse Compton photons --- at such distant and unconnected distances, whose rebirth and alignment correspond to the observed TeV gamma-ray signatures. A quite similar process, capable of solving the puzzle, could in principle also be caused by the photodisintegration of light nuclei.

\subsection{Tens of TeV photons scattering on an external gas cloud: a realistic alternative?}
 In a different, very fine tuned  model, we may consider the possible presence in the same W50 nebula around SS433 of a  very peculiar mass density in the shell gas. There the tens or hundred TeV gamma beams originated directly from SS433 may hit, split into electrons and re-emit again as TeV photons by inverse Comption scattering (ICS), as in the observed gamma beam. Such a  peculiar target may act as an optimal beam dump scattering for several tens or hundred TeV photons ejected in a past SS433 highest jet event.
 Future observations by HAWC,  H.E.S.S. and LHAASO could be able to disentangle, validate or reject each of the two models. However, the presence of such bright TeV radiation all along tens or hundred light-years distances from SS433 makes this tuned nebula mass density  (not too much dense to not be absorbed, nor too much diluted to be not interacting) an ad-hoc model somehow prone to criticisms.

\section{SS433 microquasar, precessing Jets,  GRBs and  UHECR  clustering}
We have already mentioned how micro-quasars are binary systems where a NS of a few solar masses, or an heavier BH of several or tens solar masses, are bound by gravity with an orbiting star of a  comparable or larger mass, while capturing the star mass by tidal forces. 
As the star's mass tail collapses onto the SN (or BH), the same mass fuels an accretion disk around the SN (or BH). This rotating disk typically induces asymmetric charge flows and consequently enormous currents that create a powerful toroidal magnetic field. 

The magnetic fields can undergo sudden and repetitive rapid temporal variations. These events also induce extreme pulsed spiral electric fields that accelerate charges along the disk's edge both hadrons and electron pairs. The electron at the outer edge of such disks can shine in a twin ring, as, for example, in the twin disk observed in the Crab Nebula. 
The inner free charges (protons, nucleons and electrons) spiral above and below the same accretion disk.  
These ultra-high-energy charges are quickly constrained and reduced by the extreme magnetic fields: their narrow Larmor radius forces them to form a thin cylindrical cone along the twin poles.
Eventually, these relativistic particles, both hadrons and the leptons they attract, are  ejected vertically in twin cones, aligned along the North or South axes of the NS or BH magnetic poles. 

The vibrations of the magnetic field due to their sudden shrinking and expansion, as we mentioned, led to the ejection of charge upwards and downwards even in the collimated secondary jets of electron pairs along their ultra-relativistic hadrons.

Tidal forces among the orbital star companion and the accretion disk may drive to SS433  conical precession of the spinning jet.
These tidal events, in their initial and final mass accretion phases, can feed a thin, persistent, precessing jet, whose intensity can decay over time after the destruction or collapse of the brighter companion. 
This thin, rapidly rotating jet can rarely be observed on axis with us, like the distant and fast cosmic gamma-ray burst (GRB) or the much closer and longer-lived soft gamma-ray repeaters (SGR).

 The main processes that produce the observed gamma rays in GRBs should be the ICS  \cite{fargion1998inverse}. Leptonic component, mostly electron pairs in the jet, is also leading by synchrotron radiation to some radio, X-ray, rare optical and gamma photons. The hardest gamma signals (MeV-GeV-TeV)  by ICS had been discovered since half a century and also, very recently, in surprising details along SS433. The GRB apparent huge luminosity is connected with the thin jet beaming in axis  to us  associated with their huge (thousands-millions) relativistic Lorentz factor; their variability is due to the jet spinning and precessing beam. The very recent and rarest  long life GRB and its  unexpected (and unexplained)  repetition \cite{levan2025daylongrepeatinggrb250702bde} is just an additional    argument  of such persistent, steady precessing jet model for GRBs  (as well as SGRs) \cite{fargion2006grbs}. 
 
 The SS433 system is somehow such a near, off axis,  example here to  be considered in detail, well before the same binary system  will finally collapse into a future more dramatic tidal collapse jet observable,  if in axis, as a GRB event. The tidal collapse of NS-NS,  NS-BH  or BH-BH  (with  live accretion disks) may also lead to  very rare  (because of the very narrow beam jet  alignment)  associated GRB-GW signal \cite{Fargion:2017uwg}.     
 Finally, the huge relativistic Lorentz factor acceleration needed in tens-hundreds  PeV   range  for  a nucleon,  $10^7- 10^8$,  might be associated with a more extreme relativistic one  during  W50 event explosion nearly $20.000$ years ago.   In  that case an earliest brightest supernova event,  possibly associated with the SS433  BH birth, could be also source of  an maximal GRB like jet and a corresponding highest  relativistic Lorentz factor of  $10^9- 10^{10}$,  an energy associated, for nucleons and light nuclei  to  Ultra High Energy Cosmic Rays (UHECR) events, $ E_\text{UHECR} > 10 ^{18} \text{eV}$.  

Therefore, SS433, the nearest galactic microquasar source, could be the unexpected source of a rare observed group of $4$ UHECR multiplets, clustering events around SS433 observed in the last decades by AUGER and TA~\cite{fargion2025lightest}. 
Moreover, the possible heaviest UHECR nuclei among these, with a huge relativistic Lorentz factor of $10^9-10^{10}$, ejected in the W50 explosive event about 18,000 years ago, could also reach us with some delay and in a slightly deviated direction, due to the huge charges of the heavy nuclei and the curvature of the galactic magnetic field. It can explain the most energetic and not understood UHECR event displaced by a few tens of degrees and observed a couple of years ago by TA~\cite{telescope2023extremely,fargion2024uhecr}.

\section{The unexpected SS433 separated TeV twin jet and  GZK cut-off}
As the name suggests, microquasars are just a small scale (within a parsec fraction) system of the more famous and larger quasars. Quasars are powered by million or billion solar mass BH, hidden in galactic centers, also known as Active Galactic Nuclei (AGN). These ones are able to eject much wider, longer and harder hadron  (and associated lepton) jets, even in megaparsec sizes. As we noted, SS433 is a small rare binary system containing a supergiant star that is overflowing its Roche lobe with matter onto a nearby BH, within distances of hundreds light-seconds.
Its size is small, but its proximity to us makes it an ideal test for inspecting the jet formation. 
The LHAASO experiment, the widest gamma array km-square size, recently studied the system \cite{sudoh2020multiwavelength}. The TeV beam appearance far from the source was discovered recently by H.E.S.S. and the Cherenkov Telescope
Array. Also HAWC, another large array in Mexico, revealed the disconnected TeV beams. 
The presence and collimation in such far separated beams at far distances is
quite puzzling.  Most models require surprising shock wave re-acceleration
and unexplained re-collimation.  We shall not discuss them in this article.
Here we suggest, as mentioned,  a different possibility, that one century ago a rare  explosive flare episode in this microquasar may shine, at the same time, ultraviolet hot photons and ultra-relativistic energy (UHE) proton (or nuclei) in the jet.

These proton  photo-nuclear scattering in a hot thermal bath,  their interactions, could form  Delta $\Delta^{+}$
resonances. The consequent  $\Delta^{+}$ decay may feed PeV protons  and neutral pion as well as comparable relativistic neutrons with charged pions. Such a phenomena has an analogy in cosmic volumes and in thermal infrared big bang radiation and takes place between UHECR (Ultra High Energy Cosmic Ray) at  highest $6\cdot 10^{19}-10^{20}$  eV energy  interacting with  cosmic black body photon at  $2.7K$. This phenomenon is well known under the author initials,  \cite{greisen1966end, 1966Jeptl} as the possible
”GZK cut-off”  responsible of the maximal UHECR  energy spectra values. 
We suggest here a similar processes happening in a  much small volume around SS433,   but within a more dense hot luminous photon bath,   by  a PeV  energetic proton-UV photon scattering. 
Indeed, the same Delta resonance may decay both in relic proton or as well as in  neutron secondaries. The tens PeV  neutron  may escape undeflected as a collimated energetic neutron jet. The presence of such tens PeV neutron beam,  its not-radiating flight  path, and  its much later  separated decay  explain both TeV  beta and photon trace signals, as the corner stone of our model. 

The observed 75 (up to hundreds) ly distance of the UHE neutron decay, the electron radiation resulting from the ICS as a tens of TeV gamma-ray reemerging signal, is the mechanism that solves the enigmatic and sudden reappearing TeV tracks of SS433.

\section{Ultra High Energy neutron beta decay  by  Delta resonance}
 Let us remind that a neutron must exponentially decay within nearly 877 seconds.  Assuming the corresponding flight distance at a relativistic regime one obtains:
\begin{eqnarray}
  L_n &=& 877 \; \left(E_n/m_n \right)    \;  \text{ls}\\
  L_n &=& 75  \; \left(E_n/ 25 PeV\right) \;  \text{ly}
\end{eqnarray}
 This distance of 75 ly is the first disconnected TeV signature from SS433; its size could be twice as large, assuming double the UHE neutron energy. We first use it to set a scale for the model.
Let us recall the necessary tuned resonant condition between the $\Delta$ mass creation and the photo-nuclear interaction energy
\begin{equation}
  m_{\Delta} = \sqrt{2\cdot E_p \cdot E_{\gamma}}
\end{equation}

The known $\Delta$ mass  $1232\pm 2 $ MeV, and  the  expected $25 $  PeV proton  or neutron  energy define, as a first approximation,   a photon temperature: %

\begin{equation}
 E_{\gamma} = \frac{\left(m _{\Delta}\right)^2 }{2 \cdot E_p} = 
  \frac{30.3\; \text{eV}}{\left(E_p/(25 \; \text{PeV})\right)}
\end{equation}
However, the final neutron or proton secondary of the $\Delta$  decay should take into account the lost energy in the decay of the delta of the companion pion secondary.  Therefore, to guarantee a final  $25$ PeV  neutron energy one must assume a  $10\%$ additional primary energy leading to a tuned lower  energy photon as follow: 
 \begin{equation}
  E_{\gamma} = \frac{\left(m _{\Delta}\right)^2 }{2\cdot E_p} = 
   \frac{27.6\; \text{eV}}{\left(E_p/(27.5 \; \text{PeV})\right)} =
   \frac{3.2\; 10^5 K \cdot k_{\mathrm{B}}}{\left(E_p/(27.5 \; \text{PeV})\right)}
\end{equation}

This thermal energy is in the ultraviolet range and it is nearly  $55.4$  times  more energetic than our solar spectra. Now we need a geometrical frame to estimate the photon volume and number density  where the expected SS433  proton jet could interact and, possibly, produce delta baryon resonance. This condition is not, a priori, a successful one.  

\begin{figure}[ht]
  \centering
    \includegraphics[width=0.5\textwidth]{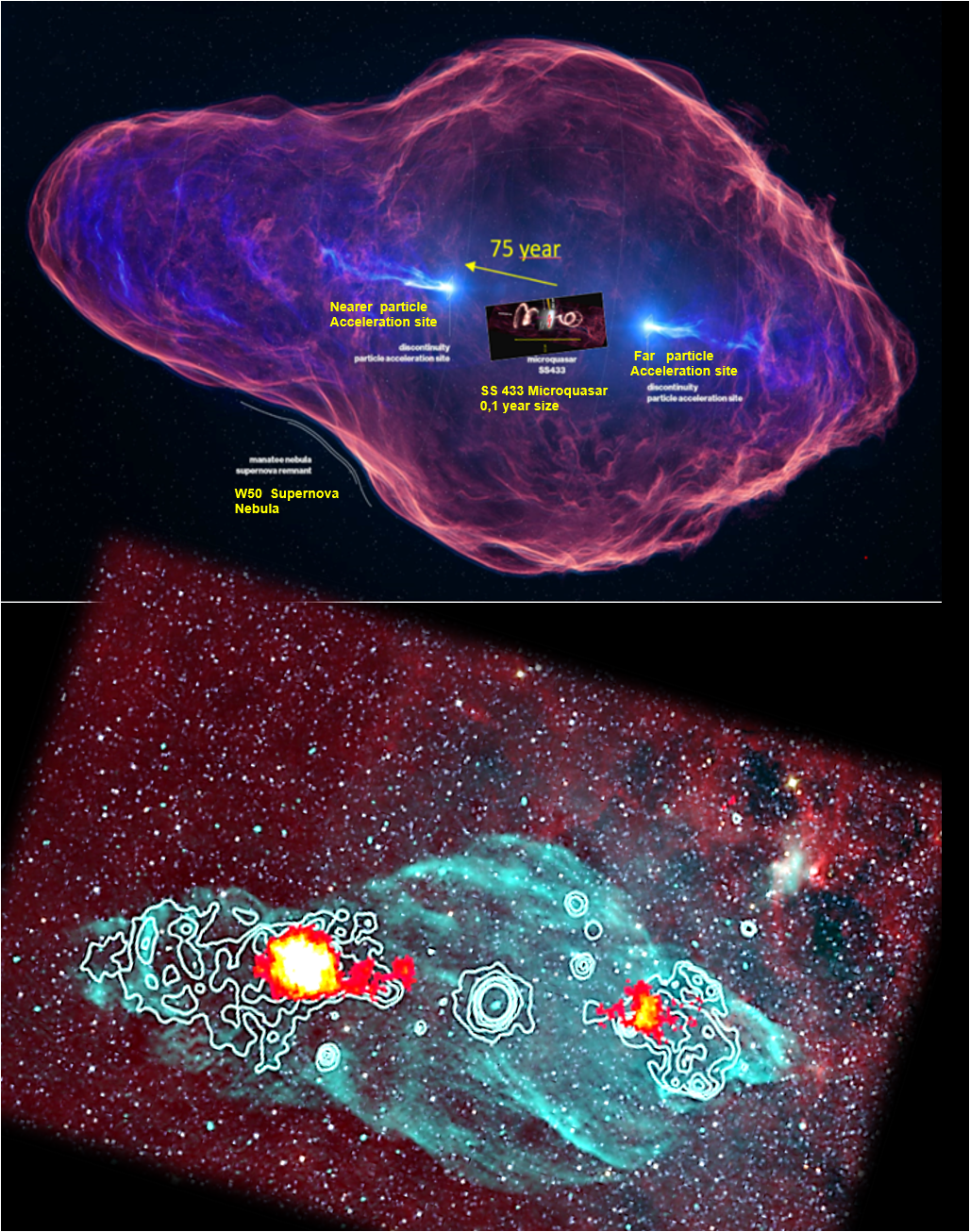} 
    \caption{
      Above an artist's impression of the SS 433 system, showing the jets (blue) and the surrounding W50 (red): the jets travel undetected for a distance of about 75 ly before suddenly reappearing as bright sources of non-thermal emission (X-rays and gamma rays) as observed by H.E.S.S., HAWC and LHAASO inside the nebula long before generated by a supernova.
      Below~\cite{hess2024acceleration} the image of the H.E.S.S. data with the additional contour of the radio radiation.
      The image of the collimated two-TeV gamma-ray beam at distances between 75 and 150 ly is very puzzling. %
     }
    \label{fig:1}
\end{figure}

\begin{figure}[ht]
  \centering
    \includegraphics[width=0.5\textwidth]{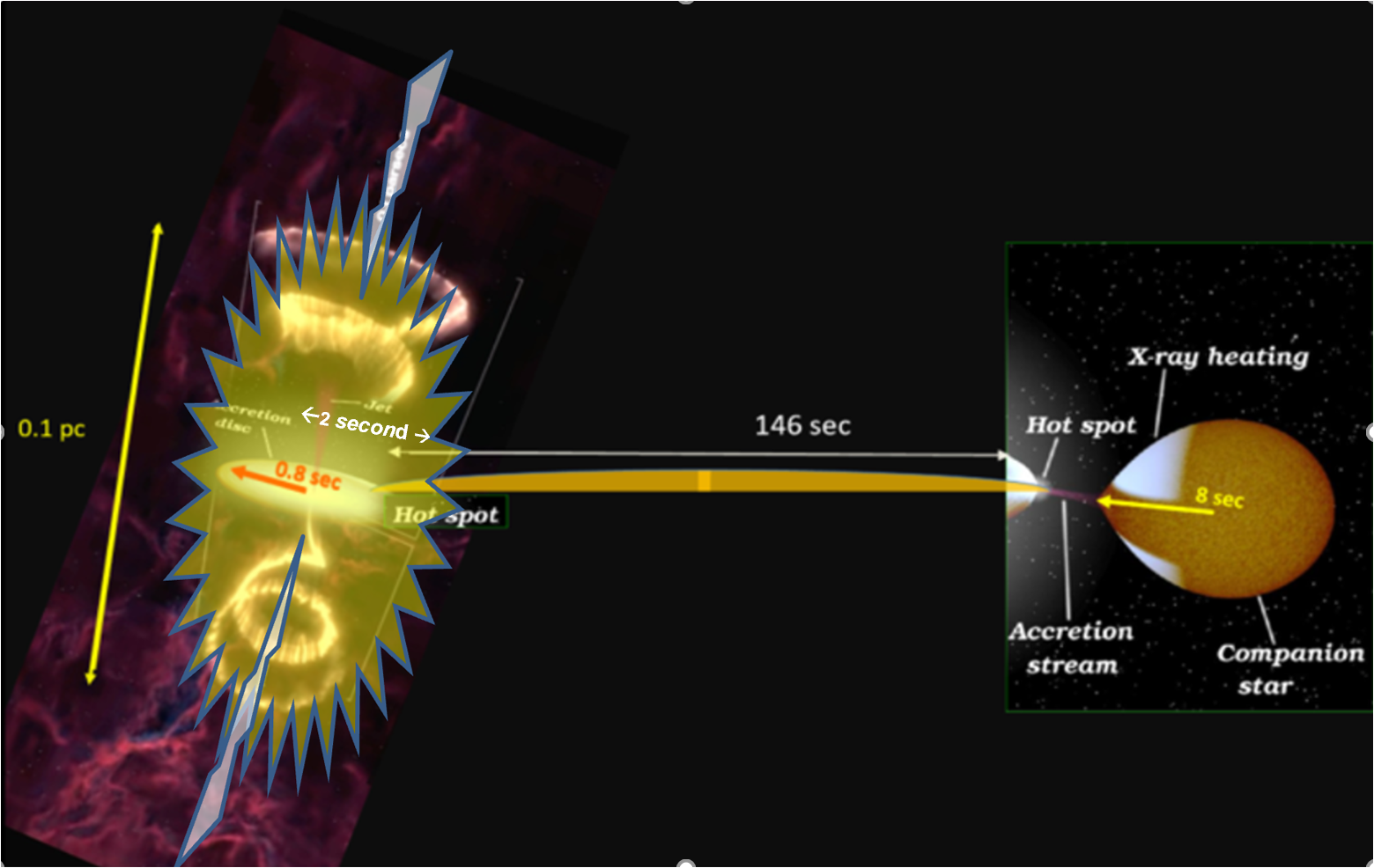} 
    \caption{A simplified description of the inner SS433 binary system, in 
    an approximated size: an accretion disk on a BH and a companion star of
    comparable mass, both of them, with a nominal ten solar mass, in their
    corresponding Kepler circular distance. The star radius and its
    particular the accretion disk, as well as the nova-like flare size  are approximated just for a  comprehensive view, but they may be  a  little  different, smaller or larger values.
    }
    \label{fig:2}
\end{figure}

 \section{Volumes and geometry for  photo-nuclear conversion into neutron}
 Let us estimate the present SS433 binary system size, as the BH companion distances, the most  probable accretion disk (whose brightening area  is the main photon volume  for the photo-nuclear event).  This estimate  may be offered,  in first approximation, by the known Kepler  orbit and by most probable BH,  and comparable  companion. 
 Their distance,  assuming a main  $M_\text{BH} = M_\text{Star}$ and $M_\text{BH}$ =10 $M_{\odot}$, and considering their observed Kepler periodicity, of 13 day period (and 162 day precession time) is:  

\begin{equation}
   D= 146 \cdot \left( \frac{M_\text{BH}}{10 \; M_{\odot}} + 
                       \frac{M_\text{Star}}{10 \; M_\odot} \right) ^{1/3} \; \text{ls}
\end{equation}

Note that such result is quite stable, because the cubic root is  not much depending on the exact system masses.   Now we may estimate as a first evaluation the photon density in such a hot flare and temperature occurring at the accretion disk area. We here assume,  as a first  estimate,  an accretion disk area comparable with our solar one.  Therefore the  Boltzmann law  imply  a  corresponding  flare luminosity  (respect  to our solar black body one) as large as the fourth power of the temperature ratio, or ${(55.4)}^{4} = 9.4 \cdot 10^{6}$,  corresponding to:

\begin{equation}
  L_\text{Flare}= 3.57 \cdot 10^ {40} \cdot \text{erg/s}
\end{equation}
This event energy (or higher ones) are comparable to a common nova flare.  The photon  number density in the same SS433  flare event is  ${(55.4)}^{3} = 1.7 \cdot 10^{5}$ the solar one  ($ n_{Th} = 3.9 \cdot 10^{12} cm^{-3}$). From the luminosity and the number density $n_{Th}$ as well as  from the following  Delta $\Delta^+$ resonant  peak  photo-nuclear resonance  cross section 

\begin{equation}
   \sigma_{\Delta} = 500 \;  \mu\text{b}  
\end{equation}
  We may derive  the probability  conversion 
\begin{equation}
    P[\gamma p \rightarrow  \Delta \rightarrow n+ \pi ] \propto   
    1-e^{- \sigma_{\Delta} n_{T} D_j}
\end{equation}

\begin{equation}
  \sigma_{\Delta}\cdot n_{Th} \cdot D_j = 23.2>>1 
\end{equation}

Where  $D_j$  is assumed to be comparable with our solar radius.
Therefore the necessary condition for an efficient proto-neutron conversion is quite well  satisfied.
It should be questioned if the longer distant neutron beam  as far as  $150$~ly is also respected. 
The needed energy will be twice larger, the  tuned  flare energy and temperature  will be twice smaller, 
and the consequent $n_{Th}$ will be $8$ times smaller, as the same photo-pion probability %
estimate that decreases  as follows: 
\begin{equation*}
    \sigma_{\Delta}\cdot n_{Th} \cdot D_j = 2.9 >1.
\end{equation*}
Therefore,  the possibility to create with a flare in SS433 able to convert a proton jet into a 27 or 54 PeV neutron jet beam is within the model possibility.

\subsection{Larmor radius for PeV proton and TeV electrons} 
There is the simple question about the proton secondaries of the Delta event considered above. 
They are charged and bent by galactic fields according to the following Larmor radius formula for the proton, $R_p$:

\begin{equation}
 R_p = (E_p/25 PeV)\cdot(B_g/(3\cdot \mu\text{G}))^{-1} \cdot 26.4 \cdot \text{ly}
\end{equation}
Therefore, they cannot rule the SS433 beam as long as  hundred years distances in a collimated way. Moreover, the comparable  tens TeV spiraling electrons,  whose re-emission  shine at TeV energies,  are also very much constrained by their Larmor radius $R_e $:

 \begin{equation}
 R_e = (E_e/25 TeV)\cdot(B_g/(3\cdot \mu\text{G}))^{-1} \cdot 0.026.4 \cdot \text{ly}
\end{equation}
These very narrow distances imply that these secondaries cannot escape far from their primary hadrons; in our model, the main PeV neutron trace.

\section{Conclusions}%

A past flare in SS433 could be the source of a 25 PeV neutron jet, possibly explaining the puzzling separated twin TeV gamma beam at 75~ly distance. The unobservable PeV neutron flight shows its presence and its beam resurgence as soon as it decays at a far distance and its electron may shine by synchrotron and ICS radiation. Its collimation is an advantage with respect to the standard model of a shock wave processes. 

An early trigger explosive event, nearly 75--150 years ago, as powerful as a nova star flare or burst, for a duration of few hours or few days in the SS433 system, may have shined almost in a fixed direction --- not along a conical precessing jet volume.
This event was around the end of War World or much earlier, around a hundred years ago. 
Nowadays, novae are observed at a rate of a ten a year in our galaxy. Such a nova event could have been escaped detection at those post war times. Indeed, the same nature of SS433 had been discovered much later, in 1977. It could be possible and worthful anyway to inspect oldest astronomical photo-plate array in that direction and at those epochs, looking for such a sudden variability signal in luminosity. The presence of such neutron PeV separated beam in SS433 may also suggest to search in other microquasars systems for disconnected signatures. 
Their statistics could define a corresponding rate of neutrino spectra in the energy range PeV to hundreds of TeV, with consequences for the ICECUBE records and neutrino flavors.
The 25 PeV neutron beta decay and its primary 27 PeV proton-pion event, while on axis toward us, may shine both of a brightest prompt ($1-2$) PeV gamma burst but also of a secondary (electron and muon) neutrino at ($0.2-0.5$) PeV.  These tuned energies are quite interesting: they define a critical energy for such neutrino events. 
The energy window could be connected to the apparent TeV-PeV energy discontinuity in the ICECUBE neutrino spectra.

\section*{Acknowledgements}
The research of D.S. was carried out in the Southern Federal University with financial support of the Ministry of Science and Higher Education of the Russian Federation (State contract GZ0110/23-10-IF).

\bibliographystyle{JHEP}
\bibliography{daf2025v26}
\end{document}